# Patterning molecular scale paramagnets at Au Surfaces: A root to Magneto-Molecular-Electronics

*Paolo Messina[1,2]

1) INSTM: Laboratory for molecular magnetism , Sesto Fiorentino, Firenze, 50119, Italy.
2)APEResearch Area Science Park, Trieste, 34012, Italy.

*Abstract* — Few examples of the exploitation of molecular magnetic properties in molecular electronics are known to date. Here we propose the realization of Self assembled monolayers (SAM) of a particular stable organic radical. This radical is meant to be used as a standard molecule on which to prove the validity of a single spin reading procedure known as ESR-STM [1]. We also discuss a range of possible applications, further than ESR-STM, of magnetic monolayers of simple purely organic magnetic molecules.

## I. INTRODUCTION

The use of the magnetic degree of freedoms in molecular electronics has been absent or restricted to a few specific cases. Transport investigations at the single molecular level are generally aimed at understanding how electrons flow through single molecules and how the single molecular transport properties can be used in the implementation of single molecular devices. Only recently the spin coupling between a single molecule and the contacts leads has proven to give Kondo effect (A typical magnetic effect arising from the coupling between the spin of an impurity an the electrons of a metal) [2]. Much work is present in the physics of quantum dots. In these system the spin of the dot is controlled by controlling the overall amount of electrons staying within the dots. Spin in these systems can be used as resource to carry information [2]. However the control of the number of electrons by charging effect requires Coulomb blockade regime. That means that the resistance between the contact leads and the dot must drastically exceed the quantum resistance. Moreover the thermal energy must be smaller than the dot charging energy resulting in very low temperatures experimental conditions. The former two constraints limit the applicability of the magnetic properties of a single DOT for technologic purposes.

Moreover a quantum dot created by lithographic techniques is still considerably bigger than a simple organic molecule. The use of molecules in place of semiconductor dots, would drastically shrink the size of the information storing unit. Also molecules offer the possibility to be self assembled at surface and do not require necessarily lithographic techniques to be organized in ordered superstructures (bottom up approach).

Purely organic molecules (radicals) offer instead the possibility to accommodate unpaired electrons coupled to each other by means of intra-molecular exchange interaction [3]. By chemical engineering, the overall spin of a molecule can be varied till large numbers (S=9..). Intra molecular ferromagnetic coupling is active also at room temperature giving rise in same cases to molecules with overall spin number S=1/2,1, 3/2, 2 at room temperature.

We needed to find nanoscale paramagnets to be ordered on a surface as probe systems on which to seek for a Radio Frequency (RF) resonance in the tunneling current flowing between the tip of an STM microscope and an underneath magnetic nanostructure ( an experiment known as ESR-STM [1]). The candidate molecule for this experiment should satisfy a number of conditions: a) being anchored on a conductive surface, b) ensuring that its magnetism is preserved, c) the intrinsic conductivity of the molecule must allow STM operation, d) the magnetism of the molecule is not suppressed during tunneling into and from its molecular orbitals, e) the overall amount of current flowing into the molecule should be sufficient to provide a detachable RF signal.

In this paper we demonstrate that a self assembled monolayer of magnetic organic radicals satisfy to the first 3 (a-c) requirements.

**THE REMAINING PART OF THIS DOCUMENT IS BEING MODIFIED.**